\documentclass{article}

     \usepackage{arxiv}

\usepackage[utf8]{inputenc} 
\usepackage[T1]{fontenc}    
\usepackage{hyperref}       
\usepackage{url}            
\usepackage{booktabs}       
\usepackage{amsfonts}       
\usepackage{nicefrac}       
\usepackage{microtype}      

\usepackage{subeqnar}
\usepackage{amsmath}
\usepackage{overpic}
\usepackage{makecell}
\usepackage{comment}
\DeclareMathOperator{\sech}{sech}
\DeclareMathOperator{\ODESolve}{ODESolve}

\title{Data-Driven Modeling of Nonlinear Traveling Waves}

\author{%
  James V. Koch \\
  Oden Institute for Computational Engineering and Sciences\\
  University of Texas at Austin\\
  Austin, TX 78712 \\
  \texttt{james.koch@austin.utexas.edu} \\

}

\begin{document}

\maketitle

\begin{abstract}
Presented is a data-driven Machine Learning (ML) framework for the identification and modeling of traveling wave spatiotemporal dynamics. The presented framework is based on the steadily-propagating traveling wave ansatz, $u(x,t) = U(\xi=x - ct + a)$. For known evolution equations, this coordinate transformation reduces governing partial differential equations (PDEs) to a set of coupled ordinary differential equations (ODEs) in the traveling wave coordinate $\xi$. Although traveling waves are readily observed in many physical systems, the underlying governing equations may be unknown. For these instances, the traveling wave ODEs can be (i) identified in an interpretable manner through an implementation of sparse regression techniques or (ii) modeled empirically with neural ODEs. Presented are these methods applied to several physical systems that admit traveling waves. Examples include traveling wave fronts, pulses, and wavetrains restricted to one-wave wave propagation in a single spatial dimension.
\end{abstract}

\section{Introduction}\label{sec:introduction}
The traveling wave is a fundamental structure that arises in many physical systems governed by partial differential equations (PDEs). Such structures are readily observed in fluid mechanics \cite{holmes2012turbulence}, condensed-matter physics \cite{Greiner2002}, optics \cite{Kutz2006}, neuroscience \cite{Bressloff2000}, and biology \cite{akhmediev2008dissipative,Feng2008}, among many other fields \cite{Yazaki1998,Mandel2008}. The physics governing such systems are often complex, multi-component, multi-scale, and nonlinear; analysis of traveling waves is often limited to direct numerical simulation of fundamental physical laws as described by PDEs. However, for the steadily propagating wave - i.e., traveling with a constant velocity and shape - the \textit{traveling wave ansatz} ($\xi = x - ct + a$) has long been used as an effective coordinate transformation that recasts the spatio-temporal dynamics of the governing PDE as a dynamical system in the single traveling wave coordinate $\xi$. The resulting dynamical system is often more mathematically tractable and readily analyzed using standard ordinary differential equation (ODE) techniques. 

In this work, proposed is methodology that aims to approximate traveling wave dynamics in the traveling wave coordinate $\xi$ - the latent space for such systems. This domain-specific knowledge is leveraged to define the inverse problem of extracting traveling wave models from data. The proposed methods are constructed around the central assumptions that (i) there exist traveling wave solutions to the governing physics, and (ii) there exists a steady velocity for which such traveling waves are viable. While observations of traveling waves satisfy the first assumption, acknowledged is that for some conditions and systems, waves may travel at non-constant velocities. Nevertheless, one can still seek a \textit{surrogate} dynamical system whose steady waveforms inherit important properties of the original system. In this article, presented are the applications of these ideas to representative PDEs that admit traveling wave solutions. In addition to reproducing the waveform, the resulting model systems are evaluated on their ability to reproduce the phase space in the neighborhood of the waveform trajectories. Where applicable, this includes the evaluation of the location, type, and stability of fixed points and location of system nullclines.

\section{Background}\label{sec:background}

\subsection{ROMs and Symmetries}
The proliferation of data has allowed for the recent rapid development of techniques and algorithms aimed at data-driven system identification and modeling \cite{brunton2019data}. In many physics problems, the spatiotemporal evolution of quantities of interest can be expressed by one or more coupled PDEs which, if known, can rarely be solved analytically. A topical example is fluid dynamics as governed by the Navier-Stokes equations, where nonlinearities and a multitude of varying scales interact to form the not only the complex flow fields associated with turbulence, but also coherent structures such as vortices, traveling waves, and shocks \cite{holmes2012turbulence}. A typical workflow for modeling such systems is discretizing the PDE over the spatial domain of interest to create a high-dimensional coupled ODE that can be integrated in time. While such methods are robust and have been used with great success, they are often computationally expensive. This is especially true in fluid dynamic simulations where the interactions of scales contribute to the overall evolution of a flowfield; thus, all scales need to be properly resolved to provide an accurate solution.

Projection-based model order reduction techniques (those based on the \textit{Singular Value Decomposition} (SVD)) seek to leverage underlying structure and patterns in the solution space of the high-dimensional full order model to perform future state predictions at low computational cost. For oscillatory flows, techniques such as the \textit{Proper Orthogonal Decomposition} (utilizing the SVD) have been successfully employed to extract the dominant space-time correlated structures \cite{Berkooz1993}. Furthermore, these structures, or modes, can be used as a reduced basis set into which the governing PDE can be projected to create a ROM \cite{brunton2019data}. POD-Galerkin is one such ROM framework that utilizes a POD reduced basis \cite{Couplet2005}.

Standard projection-based model order reduction techniques for transport-dominated physics often suffer from a non-negligible slow modal energy decay associated with the POD \cite{Reiss2018,brunton2019data}. This results in a large number of modes required to reproduce the physics. The failure of such methods is rooted in the inability of the SVD to handle translational and other symmetries (such as rotations) of the spatio-temporal field. Embedded within the POD is the assumption of separation of variables, where space and time can be cleanly decoupled. For many circumstances, this assumption may approximately hold for a reasonable time horizon. A traveling wave cannot be separated into time and spatial components; to construct a projection-based ROM without removing the translational symmetry results in a artificially high-order model. 

Removing the translational symmetry, or ``freezing'' the traveling wave, is a straightforward approach to preconditioning input data such that projection-based methods can be applied with greater success. Methods such as the \textit{Shifted Proper Orthogonal Decomposition} (sPOD \cite{Reiss2018}) and \textit{Unsupervised Traveling Wave Identification with Shifting and Truncation} (UnTWIST \cite{Mendible2020}) aim to remove the translational symmetry by offsetting the drift of traveling waves such that they appear ``frozen'' in a new reference frame. For a steadily propagating wave, by removing the translational symmetry, the first POD mode corresponds to the waveform which is invariant in time. Thus, the waveform becomes the reduced basis upon which a ROM can be built. While these methods are extremely effective in future state predictions, they do not generalize well nor offer physical insights into a given problem. Indeed, for a steadily propagating wave, any future state can be obtained by shifting the initial condition by the speed of the wave multiplied with the elapsed time. This is the linear representation of the wave, even though the physics may be inherently nonlinear. Furthermore, the applicability of these methods is dependent on the scope of the parameter space in which the training data was generated. For non-constant wave speeds and wave interactions, methods like UnTWIST provide a robust means of disambiguating between waves and extracting their trajectories, thereby providing an unsupervised method of ``freezing'' the waves. Projection methods can then be applied in the non-constant reference frame. For wave interactions and instabilities, this may yield interpretable results regarding the waves' nonlinear dynamics.

\begin{figure*}[]
        \centering
        \begin{overpic}[width=1.0\columnwidth]{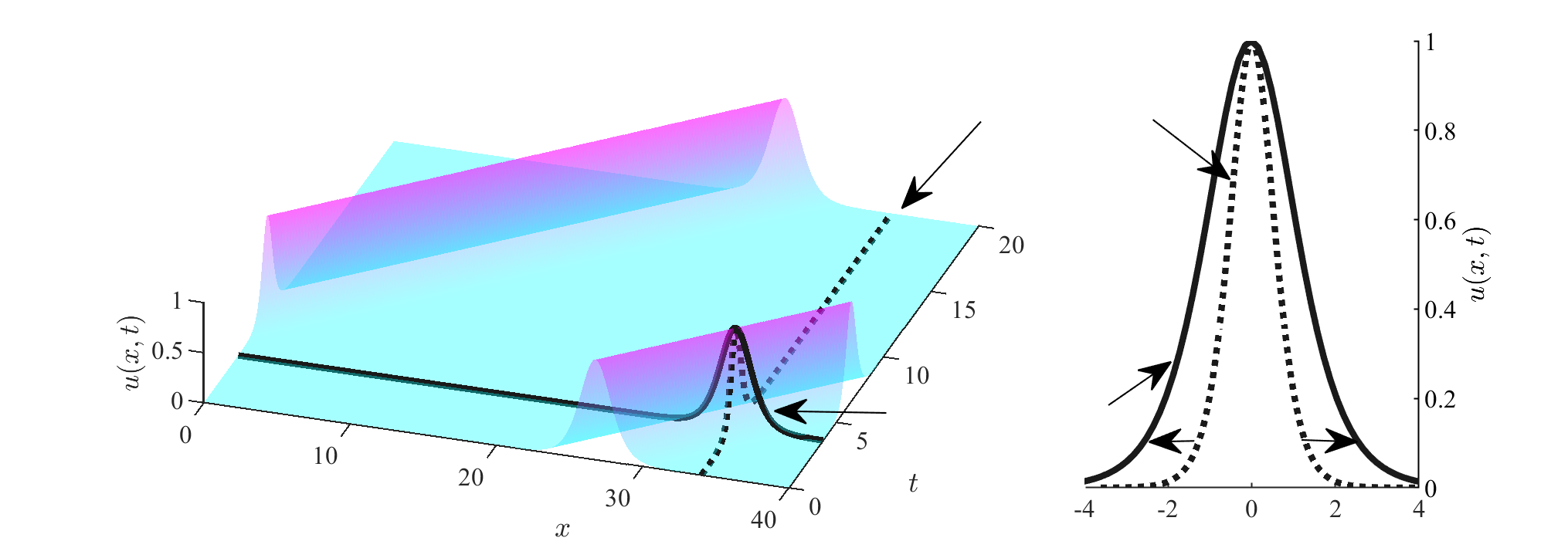}   
        \put(12,32){(a)}
        \put(70,32){(b)}
        \put(59,28){Temporal evolution}
        \put(57,8.0){Spatial evolution}
        \put(76.5,4.75){Dilation}
	    \end{overpic}  
	    \caption{A pictorial representation of the traveling wave anzatz. In (a), a KdV $sech$-pulse soliton ($c=2$) travels on a periodic boundary. In (b), a temporal and spatial slice are shown such that their peaks are centered at $t=0$ and $x=0$. For steadily propagating waves, the temporal and spatial evolution of $u(x,t)$ differ only by dilation of the temporal axis by the wave speed. Thus, for the profiles shown in (b), dilating the temporal evolution from $t$ to $ct$ (where $c=2$ for this simulation) exactly recovers the spatial waveform. }
		\label{fig:equivalency}
\end{figure*}
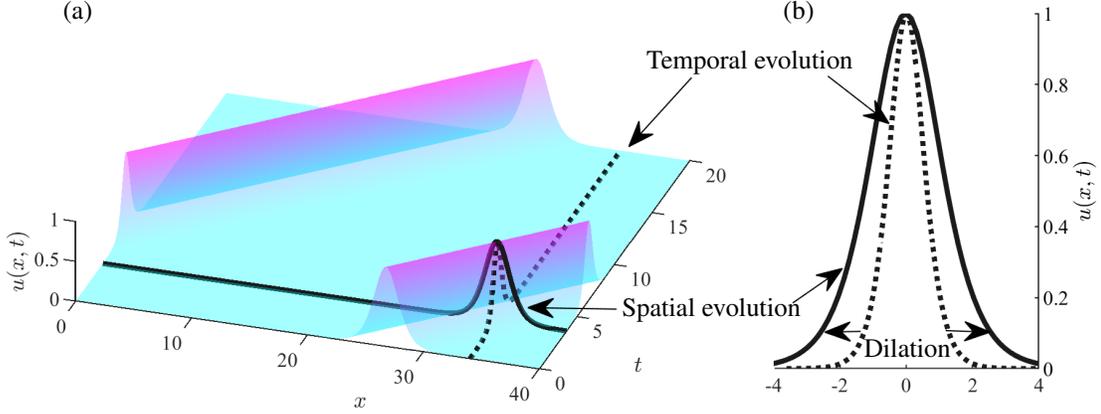

\subsection{Data-Driven System Discovery and Surrogate Modeling}
The abundance of data can also allow for the `reverse engineering' of governing physics in the form of ordinary or partial differential equations. These tasks usually aim to create either (i) parsimonious and interpretable evolution equations, or (ii) approximations to the state operator(s), a surrogate model, which can then be integrated in time with standard techniques. Methods exist for both model discovery and surrogate modeling of dynamical systems described by ODEs as well as PDEs. 

\textit{Sparse Identification of Nonlinear Dynamics} (SINDy \cite{Brunton2016}) performs a least-squares fit of data to a dynamical system constructed from a library of candidate functions, from which a sparsity-promoting algorithm chooses a parsimonious representation of the dynamics. This is a particularly attractive method because the resulting model is easily interpreted and evaluated. Should the model indeed capture the physics correctly, it is by construction generalizable and can be parameterized. However, the quality of the resultant model is strongly dependent on the quantity, quality (presence of noise, especially after differentiation), and coordinate representation of the input data. 

\textit{Neural ODEs} (NODEs \cite{NIPS2018_7892,Rackauckas2020}) are a relatively new machine learning technique for approximating the right hand side of a differential equation with a neural network. NODEs, while not as interpretable as a successful implementation of SINDy, still is a valuable method that can learn nonlinear interactions that are embedded within the input data without any \textit{a priori} knowledge of the underlying system. These differ from standard time-series forecasting techniques (deep neural networks, residual networks, long short-term memory networks, etc.) in that the neural network is integrated in time with standard ODE solvers; during training, residuals are backpropagated through both the network and the integrator. Thus the surrogate model is the inseparable unit of the ODE solver and the neural network. 

SINDy and NODEs are only two of many methods for dynamical system identification and modeling for systems described by ODEs. For PDEs, extensions of these methods have been developed and used with varying degrees of success. PDE-FIND \cite{Rudy2017} exploits the same sparsity-promotion ideas used in SINDy, though extended to include differential operators. Similarly, NODEs have been used in conjunction with convolutional neural networks to model separately the temporal and spatial evolution of a quantity of interest \cite{Rackauckas2020}. Leveraging pattern recognition techniques from the broader ML community, \textit{Physics Informed Neural Networks} (PINN) have been used extensively to perform data-driven discovery of governing equations \cite{Raissi2019}. \textit{Hidden Physics Models} \cite{Raissi2018} aim to learn governing equations from small data sets using Gaussian processes. These methods excel at identifying governing equations, though some may be more appropriate than others depending on quality, quantity, and orientation of available data.

\subsection{Traveling Wave Dynamical Systems}

The one-dimensional (1-D) \textit{Traveling Wave Ansatz} (TWA) can be stated as seeking traveling wave solutions to a governing PDE via the transformation into the traveling wave coordinate $\xi = x-ct+a$, where $c$ is the speed of an assumed right-running wave and $a$ is a spatial offset. Through this transformation, a spatially ($x$) and temporally ($t$) varying quantity of interest $u$ is recast as $u(x,t) = U(\xi = x - ct + a)$. Should the governing physics be known in the form of a PDE, the TWA is a powerful tool for recasting the governing equations as a coupled set of ODEs. By the chain rule, substitution into a governing PDE and differentiation is straightforward:
\begin{equation} \label{eq:diff1}
U_t = \frac{d U}{d \xi} \frac{d \xi}{d t} = -c U' ,
\end{equation}
\begin{equation} \label{eq:diff2}
U_{tt} =...= c^2 U'' ,
\end{equation}
where the prime denotes differentiation with respect to $\xi$. Similarly, for spatial derivatives:
\begin{equation} \label{eq:diff3}
U_x = \frac{d U}{d \xi} \frac{d \xi}{d x} = U' ,
\end{equation}
\begin{equation} \label{eq:diff4}
U_{xx} =...= U'' .
\end{equation}
Upon substitution into a 1-D wave equation, the resulting coupled system contains only ordinary derivatives with respect to the traveling wave coordinate $\xi$ (as in Eq. \ref{eq:model}) and are thus amenable to standard ODE analysis techniques.

Central to the reduction of wave equations to ODEs is the relationship between spatial and temporal evolution for steadily propagating waves. As seen in Fig. \ref{fig:equivalency}, the evolution in space and time only differ by a proportionality constant; the speed of the wave. For example, for the Korteweg-de Vries soliton moving with speed $c=2$ in Fig. \ref{fig:equivalency}, dilating the temporal axis by the speed of the wave recovers exactly the spatial shape of the soliton. Thus, temporal and spatial derivatives are equivalent up to a simple scaling as shown in Eqs. \ref{eq:diff1}-\ref{eq:diff4}.

Orbits within the phase space of traveling wave dynamical systems, as defined by their state definition, correspond to waveforms of the governing PDE. In general, traveling waves are classified as \textit{pulses}, \textit{fronts}, or \textit{wavetrains}. In phase space, these waveforms correspond to distinct orbits.  \textit{Pulses} begin and end at the same fixed point; a homoclinic orbit (Fig. \ref{fig:kdv2}). \textit{Fronts} are heteroclinic orbits that connect two fixed points (Fig. \ref{fig:kpp}). Lastly, \textit{wavetrains} form periodic orbits within phase space; a limit cycle (Fig. \ref{fig:fhnSim}). Thus, the goal of the proposed methods is to provide an estimate of the vector field that yields the observed trajectories in the system's phase space.

\section{Methodology} \label{sec:methodology}
Sought are approximate autonomous traveling wave dynamical systems for one-way wave propagation in one spatial dimension with periodic or pseudo-infinite boundary conditions.  The focus is on nonlinear waves upon saturation; that is, after one or multiple waves have formed and have reached their long-term shape and speed. This restriction in scope is appropriate as many nonlinear traveling waves in nature are observed after initial transients and saturation have occurred. Such dynamical systems are of the form:

\begin{equation} \label{eq:model}
\frac{d}{d \xi} \mathbf{y}(\xi) = f\left(\mathbf{y}(\xi) \right); ~~~ \mathbf{y}(\xi) \in \mathbb{R}^d, ~~~ f: \mathbb{R}^d \rightarrow \mathbb{R}^d ,
\end{equation}
where $\mathbf{y}(\xi)$ is the $d$-dimensional state vector of the dynamical system evolving though the traveling wave coordinate $\xi = x-ct +a$ and $f$ is a vector-valued function to be \textit{identified} or \textit{approximated}. 

For the specific task of modeling traveling waves from data, the step of discovering the governing PDE is superficial. While discovery of the governing PDE may lend some interpretability to a system's behavior, such as its' dominant balance physics \cite{Callaham2020}, examination of governing equations at face value does little to inform the behavior of traveling waves or other coherent structures. If successfully discovered, to perform analysis tasks one must either (i) simulate the PDE, or (ii) seek traveling wave solutions. In this study, the step of discovering the governing PDE is skipped. Instead, ODE system identification and modeling methods are used to directly extract traveling wave dynamical systems from data.

With the goal of producing a model of the form in Eq. \ref{eq:model}, the methodology is presented in three sections: (i) the construction of the state of the observed system, (ii) the coordinate transformation of system observations, and (ii) the regression techniques. 

\subsection{State Construction}
Assumptions regarding the mechanism(s) of wave propagation must be made to properly construct the state $\mathbf{y}$ and constrain the model to be a set of coupled first-order ODEs. Commonly encountered is diffusion-enabled wave propagation. Higher order spatial derivatives associated with diffusion or hyper-diffusion result in differentiation in $\xi$ of the same order. Therefore, if the wave propagation mechanism is assumed to be via diffusion, measurements or observations of the diffusing quantity are supplemented with their \textit{spatial} derivative. For example, for single-component ($U$, for example) diffusion, the corresponding state vector is:

\begin{equation}
    \mathbf{y} = \begin{bmatrix}
    U \\
    U'
    \end{bmatrix} = \begin{bmatrix}
    U \\
    V \end{bmatrix} .
\end{equation}
where the definition $V = U'$ ensures that the system remains first order. 

Diffusion-enabled propagation can generally be inferred by gradual transitions at the interface(s) between a wave and the quiescent medium. Exclusively advection-based propagation, usually marked by sharp fronts, discontinuities, or shocks, remains first-order after the coordinate transformation into $\xi$. More commonly encountered are advection-diffusion systems, whose measurements of the diffusing quantities would need to be supplemented with corresponding spatial derivatives. Note that successful system identification through sparse regression is not possible without correctly defining the state of the system.

\subsection{Data Transformation}

\begin{figure*}[]
        \centering
        \begin{overpic}[width=1.0\linewidth]{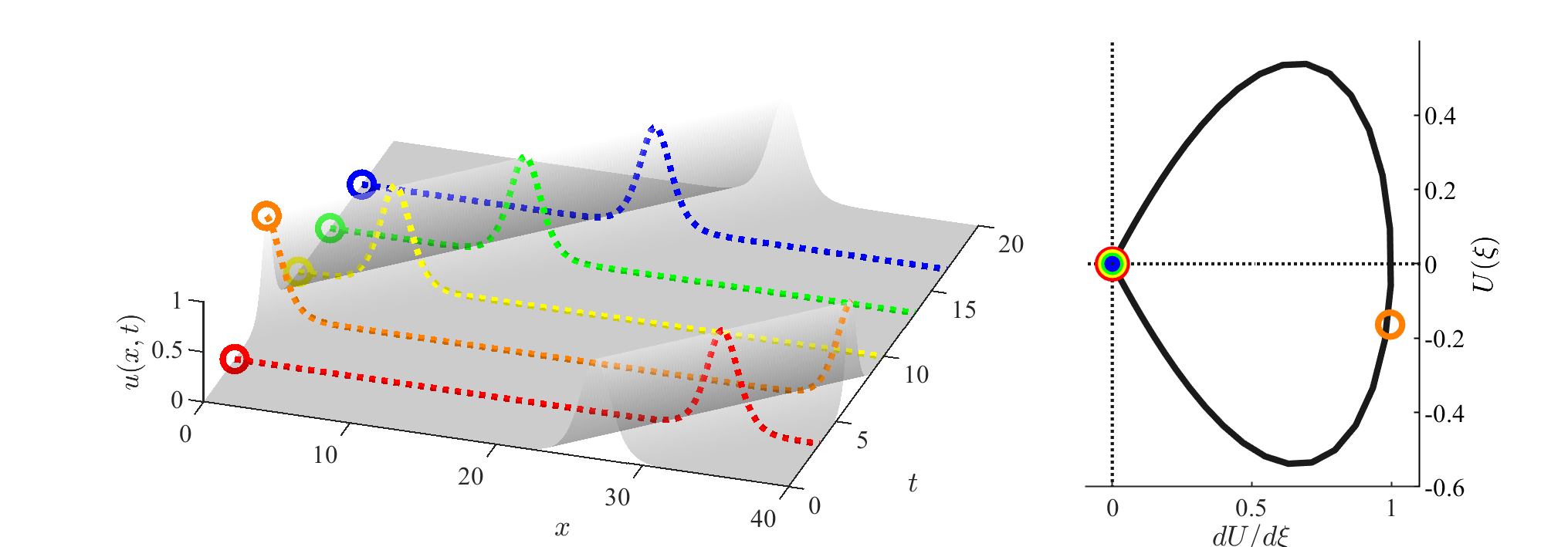}   
        \put(12,32){(a)}
        \put(70,32){(b)}
	    \end{overpic}  
	    \caption{For steadily propagating waves, snapshots of the spatiotemporal dynamics (a) map to a common trajectory through phase space (b). These distance along the common trajectory in phase space is measured by the coordinate $\xi_i = x - ct + a$. However, by re-defining $t = t - t_i = 0$ for each \textit{i}th snapshot and choosing $a = 0$, the coordinate transform becomes $\xi_i = x$. The colored circles in (a) show how values of $u(x,t)$ map to the phase plane of the system in (b). }
		\label{fig:transform}
\end{figure*}

Transforming observed spatiotemporal dynamics into the traveling wave coordinate is required to perform the proposed modeling tasks. Typically, this transformation requires knowledge of the wave speed $c$ and spatial offset $a$ such that $\xi$ is uniquely defined for an entire spatiotemporal field. However, when visualized in phase space, all system snapshots of steadily propagating waves reside on trajectories (homoclinic orbits, heteroclinic orbits, or limit cycles) described by the underlying traveling wave ODEs. Thus, only a single snapshot of the waveform is required to completely define the trajectory of the traveling wave dynamical system through phase space. This is shown explicitly in Fig. \ref{fig:transform} where multiple snapshots of a traveling KdV soliton are shown in the phase space of the system's traveling wave dynamical system. If each \textit{i}-th system snapshot is treated as the system's initial condition - the state at a newly-defined time $t_i = 0$ - the coordinate transformation can be simplified by mapping each snapshot uniquely through $(x,t_i) \rightarrow (\xi_i = x - (c \cdot t_i) + a = x + a)$. Furthermore, the spatial offset $a$ can be chosen to be zero; thus, each snapshot can be uniquely mapped to a new coordinate $(x,t_i) \rightarrow (\xi_i = x)$. This particular choice for the snapshot coordinate transformation can be exploited algorithmically in that (i) the speed of the wave need not be known and (ii) no data manipulation is required to perform the transformation.

If necessary, multiple snapshots of the traveling wave can be phase-averaged to artificially create a higher-resolution waveform (via smoothing splines or similar) or to decrease the corrupting influence of noisy measurements. For non-constant wave speeds, this coordinate transformation is approximate only; each system snapshot be unique or provide new information to the regression architecture. Generally, for these cases, more system snapshots will yield a better approximation of the system, though the minimum amount of data naturally becomes a hyperparameter to be tuned on a case-by-case basis.

Given a single system snapshot for the steady wave case, or multiple snapshots for unsteady propagation, the proposed method reduces to the procedure of performing established system identification and/or surrogate modeling techniques along the \textit{spatial} direction of spatiotemporal data (as opposed to the temporal direction). Stressed is that shifting reference frames or other input data manipulation is not necessary, nor is supplying the traveling wave speed(s) to the algorithms. 

\subsection{Regression}

The examples in Section \ref{sec:examples} use SINDy for system identification, where applicable, and neural ODEs for surrogate modeling. Stressed is that these two methods were chosen for their low barrier to entry and ease of implementation; these are not the only applicable methods nor do we claim them to be the most effective. For details regarding algorithmic implementation of SINDy we refer to Brunton et al.\cite{Brunton2016} For implementation of neural ODEs, we refer to Chen et al.\cite{NIPS2018_7892} and Rackauckas et al.\cite{Rackauckas2020} A summary of the examples provided in this paper is given in Table \ref{tab:summary}.



\section{Examples} \label{sec:examples}
\begin{table*}[ht] \label{tab:summary}
\centering
\begin{tabular}{p{1cm} p{3cm} p{5cm} p{3cm}}
 Section & System & Highlighted Property & Regression Backend \\ 
 \hline
 \hline
 \ref{sec:kdv} & KdV &  Soliton pulses & SINDy \\  \hline 
 \ref{sec:kdv_sideband} & KdV   & Sideband instability & SINDy \\   \hline
 \ref{sec:kppfisher} & KPP-Fisher   & Monotonic fronts & Neural ODEs \\   \hline
 \ref{sec:kdvb} & KdV-Burgers & Dispersive/dissipative shock fronts & Neural ODEs \\   \hline
 \ref{sec:fhn} & FitzHugh-Nagumo & Wavetrains and fast/slow dynamics & Neural ODEs \\ \hline \hline
\end{tabular}
\end{table*}

\subsection{KdV Soliton Dynamics} 

\begin{figure}[]
        \centering
        \begin{overpic}[width=.50\linewidth]{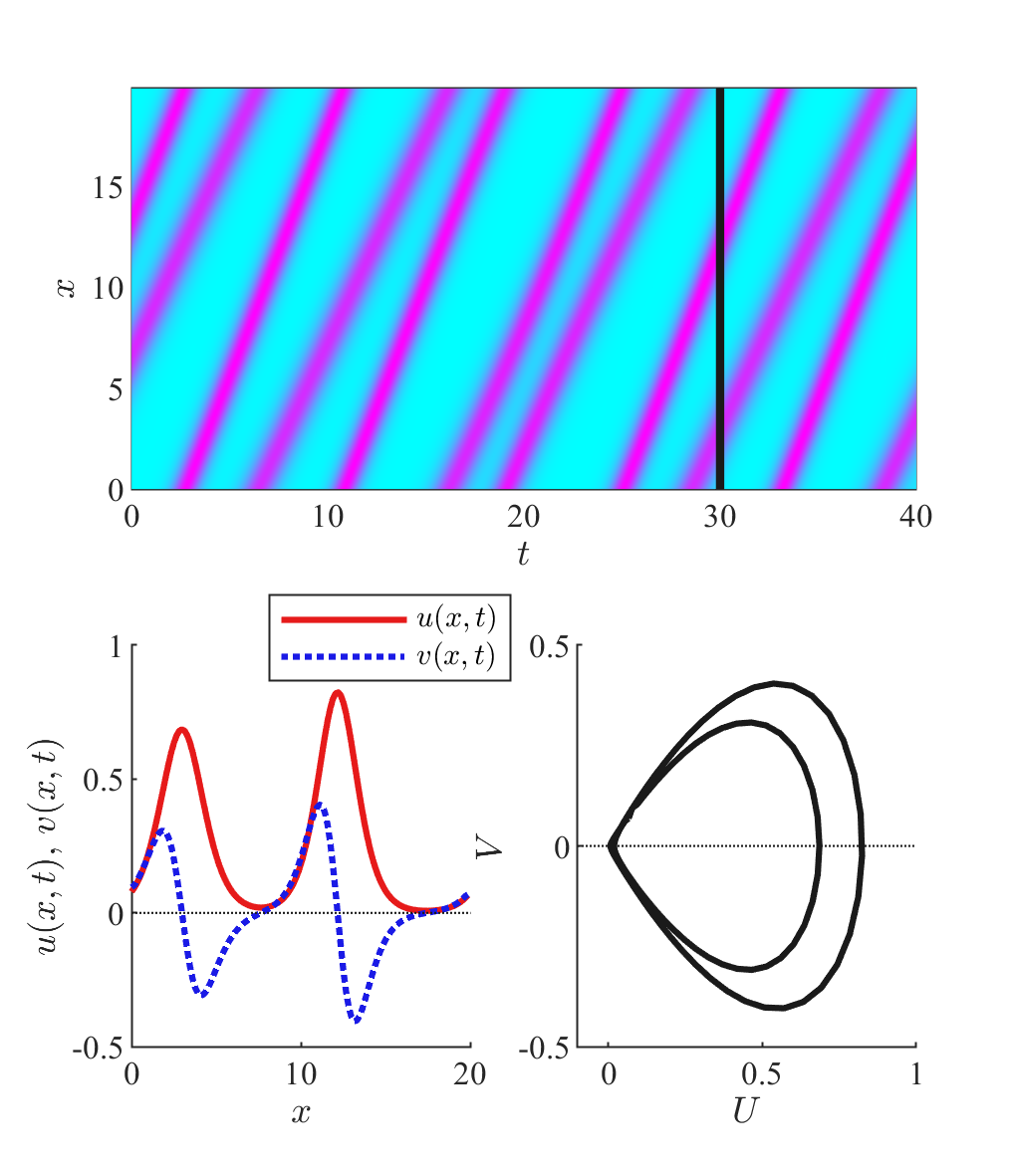} 
        \put(12,95){(a)}
        \put(12,48){(b)}
        \put(50,48){(c)}
	    \end{overpic}  
	    \caption{Two KdV $sech$-pulse solitons ($c=1.4$ and $c=1.6$) on a periodic boundary. These two solitons periodically and nonlinearly interact giving a distinctive modulational behavior. In (b), a spatial slice of the dynamics at time $t=30$ is displayed with corresponding trajectory through phase space in (c).}
		\label{fig:kdv2}
\end{figure}

The Korteweg-de Vries (KdV) equation is a historically and technically significant mathematical system that led to breakthrough discoveries in nonlinear waves. First derived for application to shallow water waves, the KdV equation's utility has since been expanded as a mathematical test bed for completely integrable systems and study of solitons. A common representation of the equation is given by:
\begin{equation} \label{eq:kdv_pde}
u_t + 6u u_{x} + u_{xxx} = 0
\end{equation}
where $u(x,t)$ is spatially $(x)$ and temporally $(t)$ variable. In the context of shallow water waves, $u$ represents the local wave height. By substituting $u(x,t) = U(\xi=x - ct + a)$ into Eq. \ref{eq:kdv_pde}, the system is transformed to the coupled set of ordinary differential equations:
\begin{equation} \label{eq:kdv_ode}
\begin{cases}

U' = V \\
V' =C_0 + cU - 3U^2 ,

\end{cases}
\end{equation}
where $c$ is the speed of the traveling wave and $C_0$ is an arbitrary constant of integration. This dynamical system has two fixed points: $(U,V) = (0,0)$ and $(U,V) = (c/3,0)$. The origin is a saddle point and the point $(c/3,0)$ is a center. The single soliton pulse is a homoclinic orbit in the $U$-$V$ plane: the beginning and end states are the origin as shown in Fig \ref{fig:transform}b. Any initial condition contained within the soliton trajectory (with the exception of the point $(c/3,0)$), results in a limit cycle in this phase space or a wavetrain in the $\xi$-coordinate.

This set of coupled ODEs can be solved analytically in terms of $\xi$, yielding the well-known \textit{sech}-pulse shape:
\begin{equation}
U(\xi) = \frac{c}{2} {sech}^2 \left( \frac{\sqrt{c}}{2} \xi \right) , 
\end{equation}
or, after back-substituting in $\xi = x - ct + a$:

\begin{equation}
u(x,t) = \frac{c}{2} {sech}^2 \left( \frac{\sqrt{c}}{2} \left(x - ct + a \right) \right) .
\end{equation}
Note that $a$ represents an arbitrary shift in $\xi$ or $x$. This waveform for $c=2$ is shown in Figs. \ref{fig:equivalency} and \ref{fig:transform}.

\subsubsection{Single Steady Soliton} \label{sec:kdv}
System identification is performed on the steady traveling wave profile of Fig. \ref{fig:equivalency}. The waveform consists of 128 grid points over the periodic domain $x \in [0,40)$. The wave propagation mechanism is assumed to be via diffusion; thus, the system state is assumed to be the coordinate $(U,U') = (U,V)$.  Treating this waveform in the typical manner as that of a time series, we apply the Sparse Identification of Nonlinear Dynamics (SINDy) algorithm to identify the dominant underlying dynamics of the flow through $\xi$. Details of the numerical simulation and SINDy parameters are given in Appendix \ref{app:kdv}. The SINDy-identified system is:

\begin{equation}\label{eq:kdv_sindy}
\begin{cases}
U' = 1.000V\\
V' = 1.996U - 2.995U^2 ,
\end{cases}
\end{equation} 
which is equivalent to Eq. \ref{eq:kdv_ode} with $C_0 = 0$ and $c=2$.

\subsubsection{Non-constant Velocities and Sideband Instabilities} \label{sec:kdv_sideband}

\begin{figure}[]
        \centering
        \begin{overpic}[width=.50\linewidth]{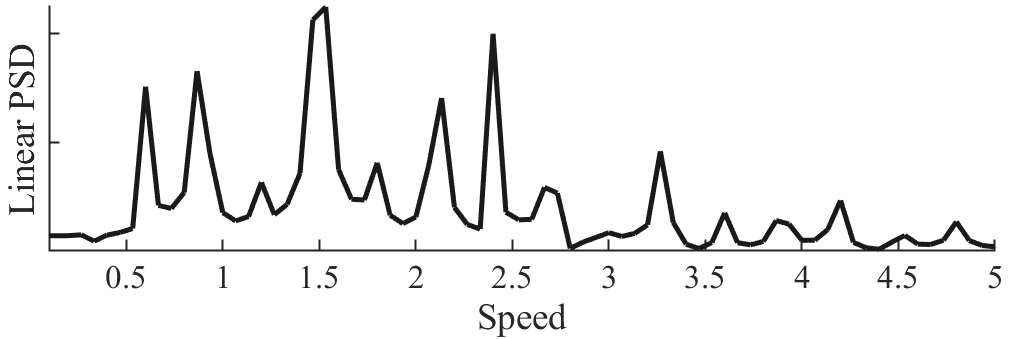}   
	    \end{overpic}  
	    \caption{Two KdV solitons of different strengths bound to a periodic domain exhibit periodic modulation that can be readily identified by sidebands present in the frequency spectrum associated with any spatial location. Sidebands are symmetric about a carrier frequency. Shown here, two solitons with speeds $c=1.4$ and $c=1.6$ yield a spectrum with carrier frequency (converted to units of speed) of approximately 1.5.}
		\label{fig:sidebands}
\end{figure}

Two solitons of speeds $c=1.4$ and $c=1.6$ constrained to a periodic domain $x \in [0,20)$ are shown in Fig. \ref{fig:kdv2}. These two solitons nonlinearly interact with one another: a phase shift is induced at each soliton collision. The oscillations in the speeds of the waves induced by this nonlinear interaction give rise to \textit{sidebands} present in the frequency spectrum associated with the temporal dynamics. The spectrum of the spatial point $u(x=0,t)$ is given in Fig. \ref{fig:sidebands} with frequency units converted into traveling wave speed. The carrier frequency is located at a speed of 1.5; this is the group velocity. Sidebands appear symmetric about the carrier frequency.

Although trajectories of this system in the $U-V$ plane are no longer unique (as evidenced by their self-intersection), one can still seek a surrogate model for the system. The goal of the surrogate model is to qualitatively reproduce the traveling waves in some representative manner; i.e., an estimate of the phase portrait for the governing physics as opposed to the specific dynamics exhibited in Fig. \ref{fig:kdv2}. In contrast the data set used for the steady soliton case, here we use the entire spatiotemporal field (as shown in Fig. \ref{fig:sidebands}) as each temporal snapshot may represent a different trajectory through phase space. After the implicit transformation from $u(x,t)$ to $U(\xi)$, SINDy is applied, yielding the dynamical system:

\begin{equation} \label{eq:kdvMod_sindy}
\begin{cases}
U' = 1.000V\\
V' = 1.371U - 2.735U^2 .
\end{cases}
\end{equation}

Equation \ref{eq:kdvMod_sindy} has the same structure and nonlinearity as that of Eqs. \ref{eq:kdv_ode} and \ref{eq:kdv_sindy}. The fixed point $(U,V) = (1.371/2.735,0) = (0.501,0)$ exists on the line connecting the coordinates of the fixed points of the separate $c=1.4$ and $c=1.6$ solitary waves. Thus the surrogate model identified by the SINDy algorithm retains representative fixed points and dynamics of the underlying system. Note, however, that this surrogate model cannot mimic the \textit{particular} trajectories as shown in Fig. \ref{fig:kdv2} without further modification (the addition of forcing terms, for example). The presented methods applied at face-value will only yield models that admit steadily propagating waves. Complete details regarding the numerical simulation and application of SINDy for this example are given in Appendix \ref{app:kdv}.

\subsection{Fisher-KPP Fronts} \label{sec:kppfisher}
The Fisher-KPP equation is a canonical mathematical system for modeling  reaction-diffusion phenomena in a variety of systems, including combustion, chemistry, and ecology \cite{fisher1937wave}. The non-dimensionalized equation reads:

\begin{equation} \label{eq:kpp}
u_t = u_{xx} + u(1-u) . 
\end{equation}

\begin{figure}[t]
        \centering
        \begin{overpic}[width=.50\linewidth]{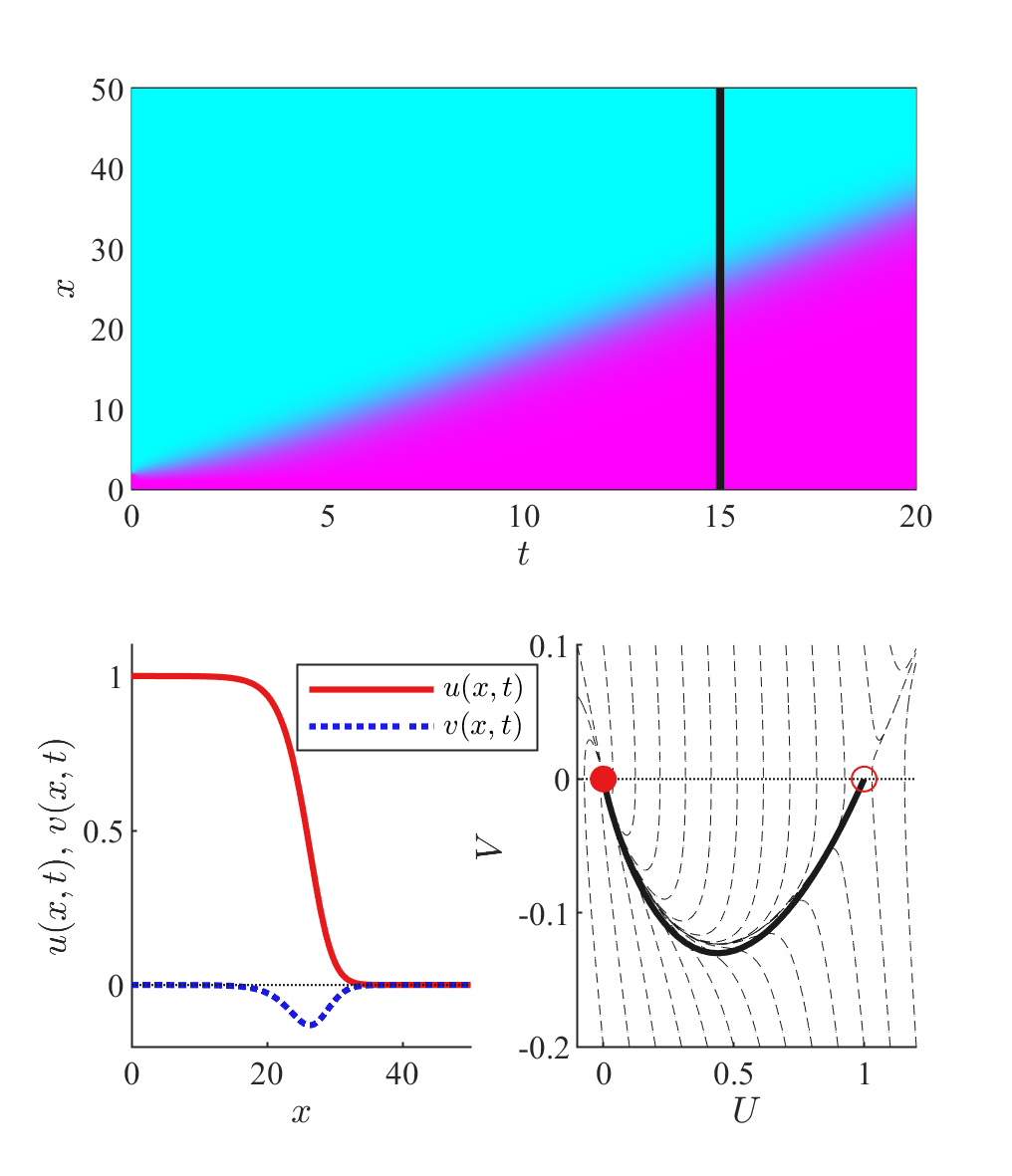}
        \put(12,95){(a)}
        \put(12,48){(b)}
        \put(50,48){(c)}
	    \end{overpic}  
	    \caption{(a) A KPP-Fisher wave front steadily propagates through the closed domain $x \in [0, 50]$. The steady waveform at time $t=15$ (the vertical cut in (a)) is profiled in (b). This single snapshot is used as the training data for the surrogate neural ODE. The front is a heteroclinic orbit in phase space, as shown in (c). }
		\label{fig:kpp}
\end{figure}

Equation \ref{eq:kpp} has steady solutions; one corresponding to $u(x,t) = 0$ and one corresponding to $u(x,t) = 1$. From inspection, one can expect values of $0<u<1$ to tend towards the $u(x,t) = 1$ state. Numerical simulation of Eq. \ref{eq:kpp} is straightforward. Figure \ref{fig:kpp} shows the developed steady traveling wave front simulated on a domain $x \in [0,50]$ with grid spacing $\Delta x = 0.5$ using the method of lines. The vertical cut shown in Fig. \ref{fig:kpp}a corresponds to the wave front profile of Fig. \ref{fig:kpp}b and phase portrait Fig. \ref{fig:kpp}c.

\begin{figure*}[t]
        \centering
        \begin{overpic}[width=1.0\linewidth]{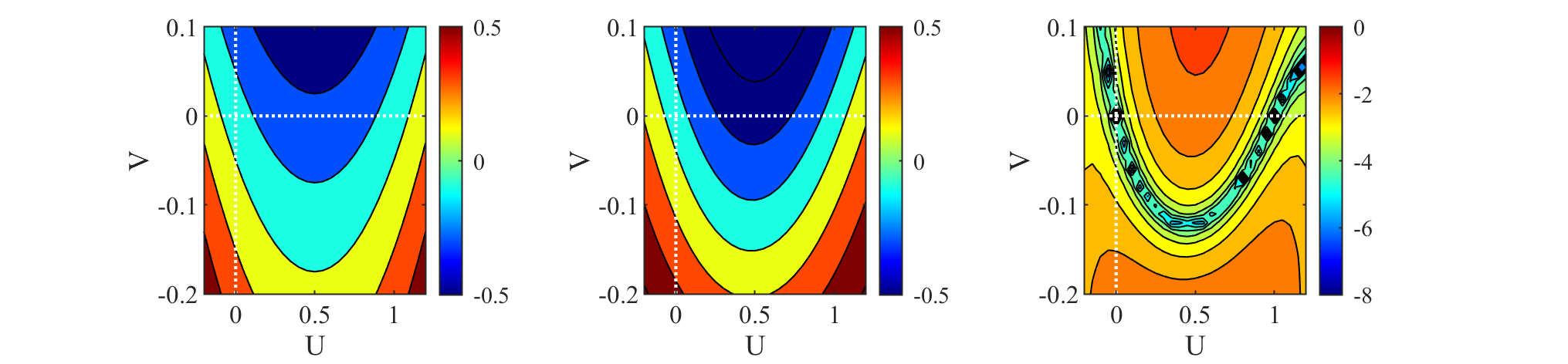}  
        \put(15,22.5){(a) Analytic}
        \put(44,22.5){(b) Model}
        \put(69,22.5){(c) L2-Error (log)}
	    \end{overpic}  
	    \caption{Visualization of the $V$-component of the vector field corresponding to the KPP-Fisher traveling wave equations and the trained surrogate neural ODE model. (a) corresponds to the analytic vector field and (b) is the trained surrogate model. The $L2$ difference between (a) and (b) is shown in (c).}
		\label{fig:kpp_eval}
\end{figure*}

Substitution of the TWA into Eq. \ref{eq:kpp} yields the coupled ODE:
\begin{equation} \label{eq:kppODE}
\begin{cases}
U' = V \\
V' = -cV - U(1-U) ,
\end{cases}
\end{equation}
where $c = 2$, the speed of the front of the wave. The traveling wave dynamical system possesses fixed points at $(U,V) = (0,0)$ and at $(U,V) = (1,0)$. Wave front solutions connect these two fixed points in a heteroclinic orbit. The $V$-component of the vector field defined in Eq. \ref{eq:kppODE} is shown in Fig. \ref{fig:kpp_eval}a.

The single trajectory shown in Fig. \ref{fig:kpp}b is used as the training data for application of the proposed method. Assuming this single trajectory is the only knowledge of the system, one can infer that the system is reaction-diffusion in type (as evidenced by diffusive appearance of  the wave front). Thus, the state is assumed to be fully defined by the coordinate $(U,U') = (U,V)$. Observations of the variable $U$ are therefore supplemented by their numerically-computed spatial derivative.  For this example, a neural ODE is trained to approximate the traveling wave dynamical system with this diffusion-enabled propagation assumption embedded by construction:
\begin{equation} \label{eq:kppNODE}
\begin{cases}
U' = V \\
V' = NN(U,V;\theta) .
\end{cases}
\end{equation}

The loss function to be minimized, $\mathcal{L}$, is the L2 error between the training trajectory (the $t=15$ snapshot) and the model trajectory, $\mathbf{z}(\xi)$, at the solution points $j$:

\begin{subeqnarray}
    \mathbf{z}(\xi) = \ODESolve(\mathbf{y}_0,NN(\theta)) \\
    \mathcal{L} = \sum_j {\left( \mathbf{y}(x_j) - \mathbf{z}(\xi_j) \right)}^2 .
\end{subeqnarray}
where $\theta$ are the neural network parameters to be optimized.

The initial condition is chosen to be a point along the wave front trajectory that is close to, but not on, the fixed point $(u,v) = (1,0)$. The network architecture used consists of one fully-connected hidden layer of size 3 that is sigmoid-activated and a linear output layer. The BFGS optimization routine was used to minimize the loss to a value below $1\cdot10^{-5}$. Figure \ref{fig:kpp_eval} shows the resultant vector field of the surrogate model compared with the analytic ground-truth. The wave front is successfully reconstructed to high accuracy. Although the surrogate model was only trained on the single steady wave front trajectory, the  model is at least first order accurate for the region of phase space shown in Fig. \ref{fig:kpp_eval}. Full details regarding the numerical simulation and neural ODE architecture are given in Appendix \ref{app:fisherkpp}.

\subsection{KdV-Burgers Dispersive-Dissipative Shock Wave} \label{sec:kdvb}
\begin{figure}[t]
        \centering
        \begin{overpic}[width=.50\linewidth]{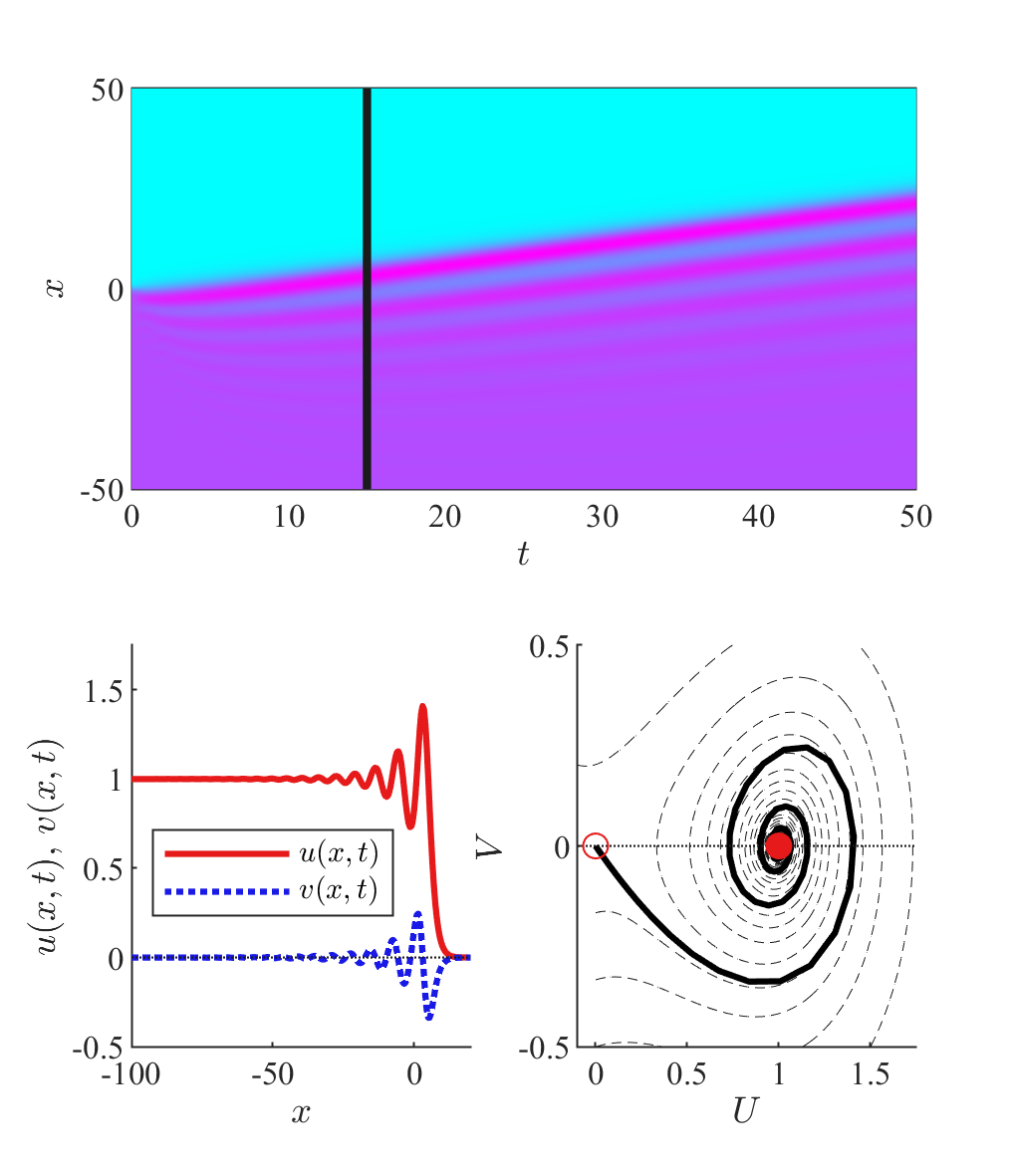}   
        \put(12,95){(a)}
        \put(12,48){(b)}
        \put(50,48){(c)}
	    \end{overpic}  
	    \caption{(a) Solution of the KdV-Burgers PDE on the closed domain $x \in [-100,300]$. (b) The 1-D waveform at time $t=15$. This waveform is the training set for the neural ODE. (c) The phase space representation of the training data waveform.}
		\label{fig:kdvb}
\end{figure}

The viscous Burgers' equation possesses a prototypical \textit{shock structure solution} connecting pre- and post-shock states that can be analytically solved for upon substitution of the TWA. Regularization with a dispersive term results in the KdV-Burgers equation\cite{johnson1970non}, given as:
\begin{equation} \label{eq:kdvb}
u_t + u u_x + u_{xxx} = \nu u_{xx} .
\end{equation}
For certain choices of viscosity ($\nu$), the KdV-Burgers equation admits a steadily propagating front connecting upstream and downstream states via an oscillatory front. This frontal structure is a prototypical example of a \textit{dispersive-dissipative} shock. Numerical integration of the PDE on the domain $x \in [-100,300]$ with $\Delta x = 0.5$ is shown in Fig. \ref{fig:kdvb}a, with the wave profile displayed in Fig. \ref{fig:kdvb}b and corresponding trjectory through phase space in Fig. \ref{eq:kdvb}c. The initial condition for the simulation is $u(x,t = 0) = \frac{1}{2}\left(1 + \tanh(x)\right)$. Viscosity is $\nu = 0.05$. Substitution of the TWA into Eq. \ref{eq:kdvb} yields the coupled ODEs:

\begin{equation} \label{eq:kdvbODE}
\begin{cases}
U' = V \\
V' = -\frac{1}{2} U (U - 1) + \nu V .
\end{cases}
\end{equation}

For the value of viscosity chosen, $\nu = 0.05$, the fixed points of the system exist at $(U,V) = (0,0)$ and $(U,V) = (1,0)$. The origin is a saddle point. The coordinate $(1,0)$ is an attracting spiral - this is clearly observed in Fig. \ref{fig:kdvb}c.

\begin{figure*}[t]
        \centering
        \begin{overpic}[width=1.0\linewidth]{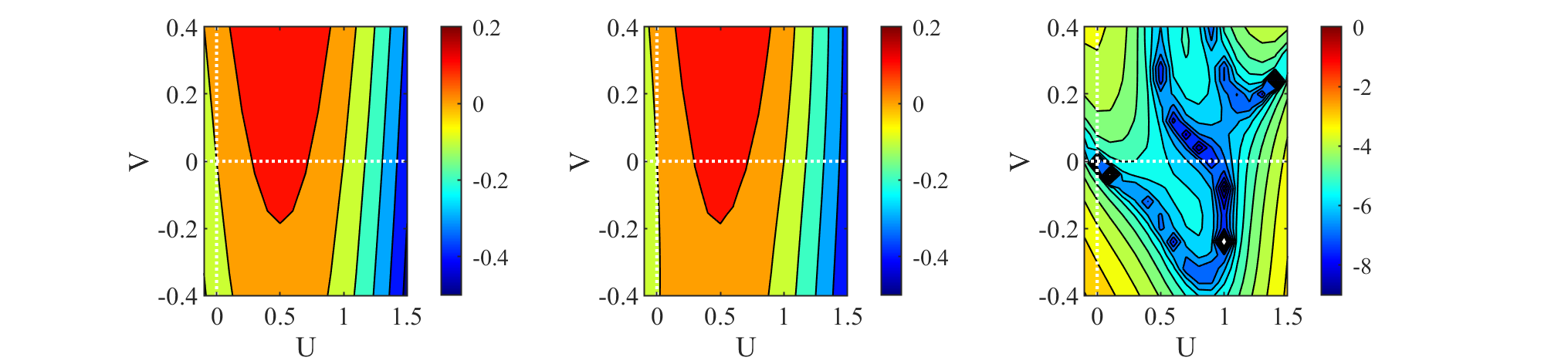} 
        \put(15,22.5){(a) Analytic}
        \put(44,22.5){(b) Model}
        \put(69,22.5){(c) L2-Error (log)}
	    \end{overpic}  
	    \caption{Visualization of the $V$-component of the vector field corresponding to the KdV-Burgers traveling wave equations and the trained surrogate neural ODE model. (a) corresponds to the analytic vector field and (b) is the trained surrogate model. The $L2$ difference between (a) and (b) is shown in (c).}
		\label{fig:kdvb_eval}
\end{figure*}

From inspection of the waveform in Fig. \ref{fig:kdvb}, the assumption of diffusion-enabled propagation is made for this system. Therefore, to properly construct the state $(U,U') = (U,V)$, measurements of the variable $U$ are supplemented with their numerically-computed spatial derivatives. For this example, a neural ODE is trained to construct a model. The diffusion-enabled propagation assumption is embedded in the same manner as in Eq. \ref{eq:kppNODE}. The neural network architecture consists of a single sigmoid-activated hidden layer of size 3 and a linear output layer. The training data is a single trajectory through phase space truncated to only be the data points not representing rest states (either of the fixed points).

Comparisons of the trained surrogate model and the analytic $V$-component of the vector field is provided in Fig. \ref{fig:kdvb_eval}. For the single trajectory training set, the surrogate model is approximately second-order accurate for the entirety of phase space shown in Fig. \ref{fig:kdvb_eval}. Full details regarding the numerical simulation and neural ODE architecture are given in Appendix \ref{app:kdvb}.

\subsection{FitzHugh-Nagumo Wavetrain} \label{sec:fhn}
The diffusive Fitz-Hugh Nagumo (FHN) model is a simple mathematical model that inherits the features of reactivity, spiking, and refractoriness from the Hodgkin-Huxley model for neuronal dynamics \cite{fitzhugh1955mathematical}. The spatially extended model is given by:
\begin{equation}\label{eq:fhn}
\begin{cases}
u_t = \nu u_{xx} - v + u(1-u)(u-a)\\
v_t = \epsilon (b u - v) ,
\end{cases}
\end{equation}
where $u(x,t)$ is the voltage in an axon and $v(x,t)$ is a slowly evolving recovery variable. In Eq. \ref{eq:fhn}, only the voltage is diffusive, though in general, both the recovery variable and the voltage may diffuse and at different rates. The dynamics of the voltage follow a cubic nonlinearity that mimics reactivity; upon reaching a threshold voltage, $a$, the reaction term $u(1-u)(u-a)$ is positive and causes a spike in the dynamics. Depending on parameter values chosen, the FHN system can exhibit intersting dynamics such as solitary pulse propagation and the development of pulse trains.

\begin{figure}[t]
        \centering
        \begin{overpic}[width=.50\linewidth]{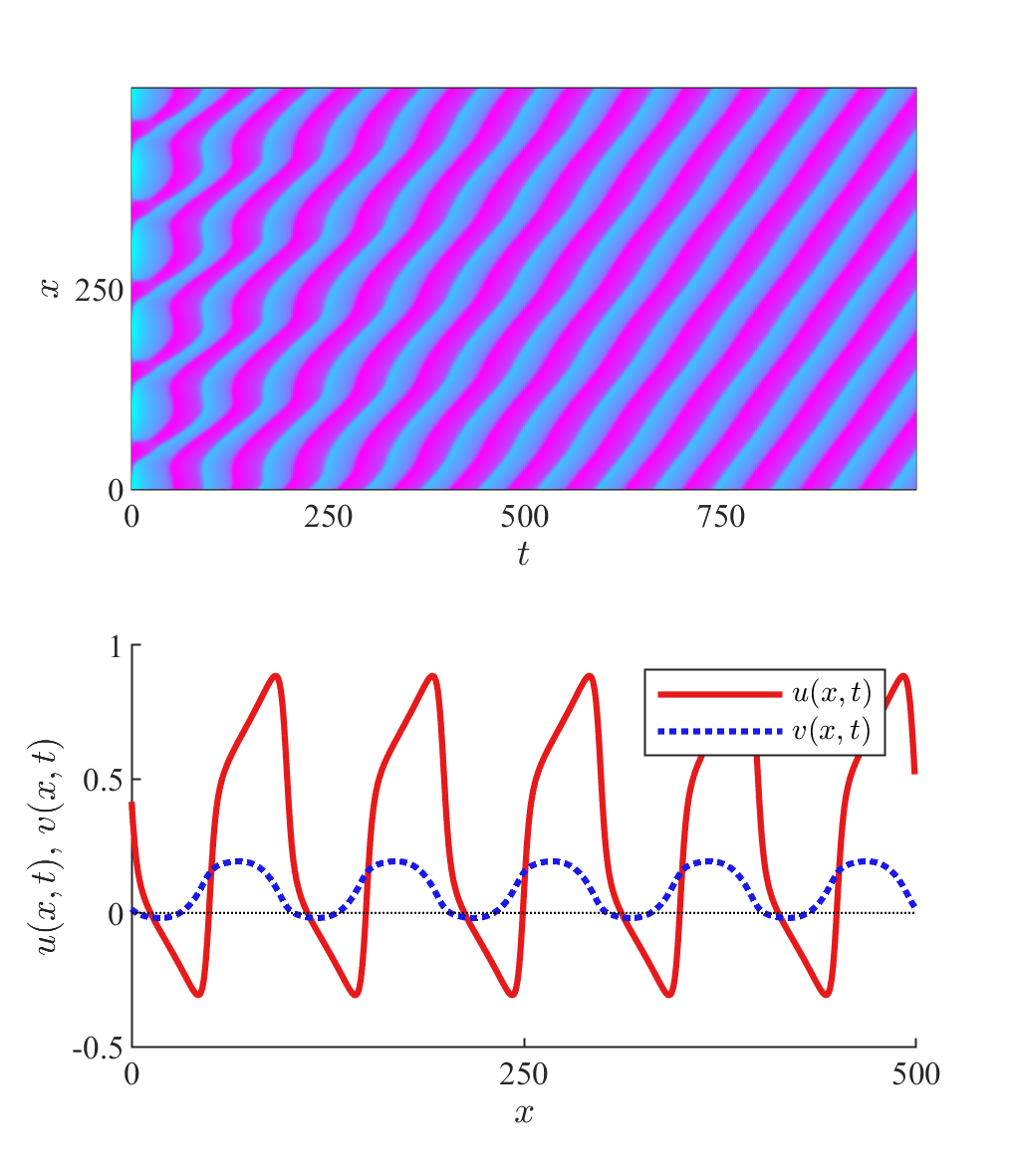}   
        \put(12,95){(a)}
        \put(12,47){(b)}
	    \end{overpic}  
	    \caption{In (a), displayed is a simulation of the FitzHugh-Nagumo equations on a periodic domain of length $L=500$ starting from a 5-peaked sinusoidal initial condition. Simulation parameters are listed in Tab. \ref{tab:fhn}. In (b), the steady-state waveform is shown.}
		\label{fig:fhnSim}
\end{figure}

For the present example, a surrogate model for a FHN pulse train is trained and analyzed with the goal of identifying the cubic nonlinearity. Numerical integration of the model in Eq. \ref{eq:fhn} was performed via the method of lines for $\epsilon = 0.05$, $\nu = 1.0$, $a = -0.1$, and $b = 0.3$ on a periodic domain of length $L = 500$. Beginning with an initial condition of $u(x,0) = \frac{1}{2} + \frac{1}{2}\sin\left(\frac{10 \pi}{L} x\right)$ and $v(x,0) = \frac{1}{2} + \frac{1}{2}\cos\left(\frac{10 \pi}{L} x\right)$, the dynamics quickly settle into a pulse train with five distinct pulses that travel at a speed of $c=1.11$. The spatiotemporal evolution of $u(x,t)$ and the steady waveform are shown in Fig. \ref{fig:fhnSim}. Substitution of the TWA into Eq. \ref{eq:fhn} yields the three coupled ODEs:
\begin{equation} \label{eq:fhnODE}
\begin{cases}
U' = W \\
V' = \frac{\epsilon}{c} \left( b U - V \right) \\
W' = \frac{1}{\nu} \left( -c W + V - U\left( 1 - U \right) \left( U - a \right) \right) .
\end{cases}
\end{equation}

\begin{figure}[t]
        \centering
        \begin{overpic}[width=.50\linewidth]{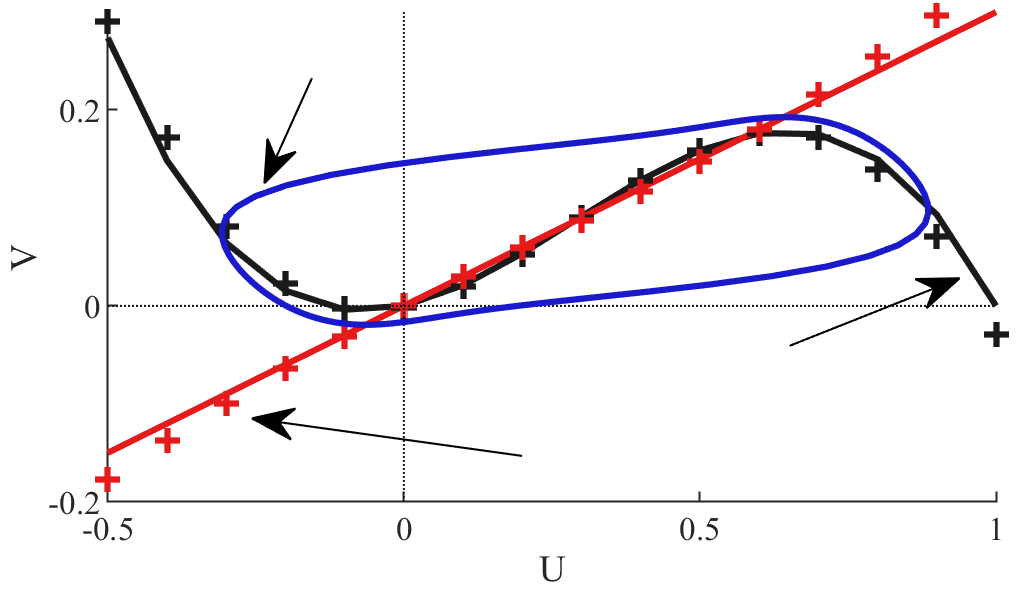}   
        \put(53,12.5){$V$-nullcline}
        \put(56,22.5){$U$-nullcline}
        \put(32,50){Projection of limit cycle}
        \put(32,45.5){on $W=0$ plane}
	    \end{overpic}  
	    \caption{$U$- and $V$-nullclines of the FHN traveling wave dynamical system and the trained surrogate neural ODE. The $U$-nullcline (black) corresponds to the cubic reaction term of the governing PDE. The $V$-nullcline (red) corresponds to the linear slow return of the recovery variable. Solid lines correspond to the analytic nullclines. Plot markers correspond to the numerically-interpolated values from the neural ODE.}
		\label{fig:fhn_null}
\end{figure}

These three-dimensional dynamics are examined on the plane $W=0$, which exposes the structure of the nonlinearity embedded in $W'$. Two nullclines exist on this plane defining where $V'=W'=0$. The nullcline $W'=0$ corresponds to the cubic polynomial describing the excitability or reactivity of the medium; $V = U(1-U)(U-a)$ (see Fig. \ref{fig:fhn_null}). 

A surrogate neural ODE is trained against the single wavetrain profile shown in Fig. \ref{fig:fhnSim}. The propagation mechanism is assumed to be via diffusion. The state of the system is therefore given by the coordinates $(U,U',V) = (U,W,V)$. The neural ODE is therefore constructed with the constraint that $U' = W$:
\begin{equation} \label{eq:fhnNODE}
\begin{cases}
U' = W \\
V' = NN_{1}(U,V,W;\theta) \\
W' = NN_{2}(U,V,W;\theta) .
\end{cases}
\end{equation}

The neural ODE is trained via \textit{minibatching} in order to avoid issues in obtaining trivial local minima. The objective function is:
\begin{equation}
    \mathcal{L} = \sum_{i=1}^{N/k} {\left( \mathbf{y}(x_{ki} + \Delta \xi) - \ODESolve(\mathbf{y}(x_{ki}),NN(\theta))\Bigr\rvert_{\xi = \Delta \xi} \right)}^2 ,
\end{equation}
where N is the total number of data points along the trajectory, $k$ is the degree of sub-sampling (every $k$-th point along the training trajectory is used to seed an ODESolve), and $\Delta \xi$ sets where along the trajectories the ODESolve is evaluated. For this example, $k=4$ and $\Delta \xi = 5$. This method exchanges the difficulty associated with periodic training data (such as arriving at trivial local minima) for a $N/k$-fold increase in ODE solves. Although this training process is undoubtedly slower, for oscillatory and stiff problems, this method of training is more robust. The neural network architecture consists of a single sigmoid-activated hidden layer of size 3 and a linear output layer. 

\begin{figure}[t]
        \centering
        \begin{overpic}[width=.50\linewidth]{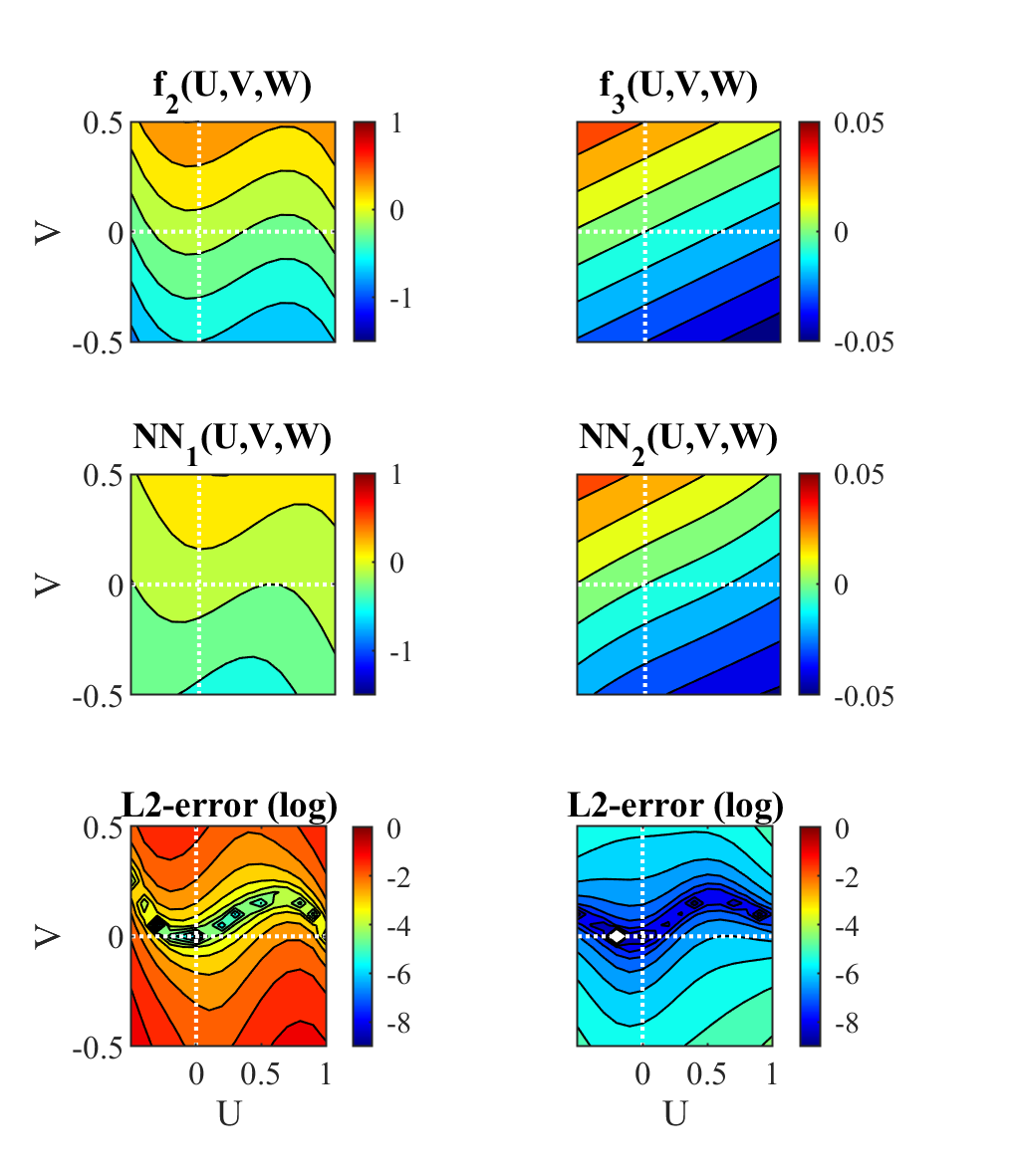}   
	    \end{overpic}  
	    \caption{Visualization of the vector field for the FHN traveling wave equations and trained surrogate neural ODE model on the plane $W=0$. The traveling wave ODE (Eq. \ref{eq:fhnODE}) second (cubic reaction nonlinearity) and third (linear recovery) components are displayed in top row, respectively. The corresponding surrogate model components are displayed in the middle row. The $L2$ difference between the analytic and model vector field components are given in the bottom row.}
		\label{fig:fhn_eval}
\end{figure}

Figure \ref{fig:fhn_eval} compares representative 2-D vector fields at the plane $W = 0$ for the surrogate model and the analytic traveling wave ODEs. The numerically-interpolated nullclines $V'=W'=0$ for both systems are shown in Fig. \ref{fig:fhn_null}. The solid lines correspond to the analytic nullclines; the plot markers correspond to the interpolated locations from the surrogate model. The analytic $W' = 0$ nullcline and the cubic polynomial fit of that of the surrogate model are:
\begin{subeqnarray}
    V_{true}(U) = -U^3 + 0.865 U^2 + 0.135 U \\
    V_{model}(U) = -1.06 U^3 + 0.907 U^2 + 0.125 U + 0.00162 .
\end{subeqnarray}
Thus, the underlying cubic linearity of the system has been successfully extracted. Full details regarding the numerical simulation and neural ODE architecture are given in Appendix \ref{app:fhn}.

\section{Conclusion} \label{sec:conclusion}
The coordinate transformation $\xi = x - ct +a$ is a powerful tool to recast wave equations as a set of ordinary differential equations. In this paper, we leveraged this co-ordinate transformation to perform system identification and surrogate modeling within this latent space of the governing equations. Algorithmically, this co-ordinate transformation is \textit{implicit}; the presented methodology reduces to applying well-established time series modeling techniques along the spatial dimension of a system snapshot. For the present study, neural ODEs and Sparse Identification of Nonlinear Dynamics are used for this task. These ideas were applied to several examples, including the soliton-producing KdV equation, fronts of the KPP-Fisher and KdV-Burgers equations, pulse trains of the FHN equation, and finally rotating detonation waves. In addition to reproducing observed waveforms to high accuracy, the resulting models captured the structure embedded within the traveling wave ODEs, including location and type of fixed points, limit cycle behavior, and heteroclinic and homoclinic orbits.

\section*{Acknowledgements} 
This work was supported by the US Air Force Center of Excellence on Multi-Fidelity Modeling of Rocket Combustor Dynamics award FA9550-17-1-0195. 

\section*{Data Availability}
The data that support the findings of this study are available from the corresponding author upon reasonable request.

\appendix
\section{Data, Model Architectures, and Parameters}

\subsection{KdV Solitons and Instabilities} \label{app:kdv}
\subsubsection{Single Soliton}
The simulation of Fig. \ref{fig:equivalency} was performed with a pseudospectral method on the periodic domain $x \in [0, 40)$ with 256 grid points. The initial condition is $u(x,t=0) = \frac{1}{2} \sech ^2 \left( \frac{\sqrt{c}}{2} \left(x - 15\right) \right)$ with $c=2$. The simulation was integrated with a fourth order Runge-Kutta integrator to time $t=20$. The snapshot matrix contains 300 time steps. PySINDy \cite{de2020pysindy} was used for the implementation of the sparse regression backend and function library. The library used contained polynomials up to order five. A Sequentially thresholded least-squares optimization routine was used with a threshold hyperparameter of 0.8. Only a single snapshot (the initial condition) was used in the regression. 

\subsubsection{Sideband Instability}
The simulation of Fig. \ref{fig:kdv2} was performed with a pseudospectral method on the periodic domain $x \in [0, 20)$ with 128 grid points. The initial condition is $u(x,t=0) = \frac{1}{2} \sech ^2 \left( \frac{\sqrt{1.4}}{2} \left(x - 6.7\right) \right) + \frac{1}{2} \sech ^2 \left( \frac{\sqrt{1.6}}{2} \left(x - 13.3\right) \right)$. The simulation was integrated with a fourth order Runge-Kutta integrator to time $t=150$. The snapshot matrix contains 3000 time steps. PySINDy was used for the sparse regression backend with a library of polynomials up to degree 5. The optimizer used was the sequentially thresholded least-squares with the threshold hyperparameter set to 0.8. All snapshots were used in the regression. 

\subsection{Fisher-KPP Fronts} \label{app:fisherkpp}

The simulation of Section \ref{sec:kppfisher} was performed via third-order finite differences in space and a fourth-order Runge-Kutta integrator in time. The 1-D domain is $x \in [0,50]$ with 201 grid points and with boundary conditions $u(0,t) = 1$ and $u(50,t) = 0$. The initial condition is $u(x,t=0) = \mathcal{H}\left(2-x\right)$ where $\mathcal{H}(\cdot)$ is the Heaviside step function. The system is integrated to time $t=20$. The snapshot matrix contains 101 snapshots (including the initial condition). 

The neural ODE surrogate model is comprised of the coupled ODE solver and feed-forward neural network. The ODE solver used is a fifth-order Runge-Kutta integrator with fourth-order interpolant. The neural network is comprised of an input layer of dimension two, one fully-connected hidden layers of dimension 3 (sigmoid-activated), and an output layer of dimension two (linear). A single snapshot of the wave front is used for training the neural ODE. The loss function to be minimized, $\mathcal{L}$, is the L2 error between the training trajectory ($t=15$ snapshot) and the model trajectory, $z(\xi)$:

\begin{subeqnarray}
    \mathbf{z}(\xi) = \ODESolve(\mathbf{y}_0,NN(\theta)) \\
    \mathcal{L} = \sum_i {\left( \mathbf{y}(x_i) - \mathbf{z}(\xi_i) \right)}^2 .
\end{subeqnarray}
Because wave fronts connect two fixed points in a homoclinic orbit, initializing the neural ODE at the ``upstream'' rest state will cause an optimization failure. The initial condition is therefore chosen to be a location in phase space along the homoclinic orbit and near, but not at, the ``upstream'' rest state. A total of 75 data points (representing the front exclusively) are used in training. The BFGS optimization routine is used to select the neural network parameters $\theta$ subject to the loss function. The convergence criteria is arbitrarily set to obtaining a loss less than $10^{-5}$.

\subsection{KdV-Burgers Fronts} \label{app:kdvb}

The simulation of Section \ref{sec:kdvb} was performed via third-order finite differences in space and a fourth-order Runge-Kutta integrator in time. The 1-D domain is $x \in [-100, 300]$ with 801 grid points and boundary conditions $u(-100,t) = 1$ and $u(300,t) = 0$. The simulation initial condition is $u(x,t=0) = \frac{1}{2} \left(1 - \tanh(x)\right)$. The system is integrated to time $t=50$. The snapshot matrix contains 51 snapshots (including the initial condition). 

The surrogate model architecture and loss function are identical to those of the Fisher-KPP system. The neural ODE surrogate model uses a fifth-order Runge-Kutta integrator with fourth-order interpolant for the ODE solver. The neural network is comprised of an input layer of dimension two, one fully-connected hidden layers of dimension 3 (sigmoid-activated), and an output layer of dimension two (linear). A single snapshot of the wave front is used for training the neural ODE. Similar to Fisher-KPP fronts, the neural ODE is initialized at a location along the homoclinic orbit connecting the system's two fixed points. A total of 256 data points (corresponding to a length of $\xi = 128$) are used in training. This represents the frontal dynamics only; excluded are the steady-state regions. The loss function to be minimized is the L2 error between the training trajectory ($t=15$ snapshot) and the modeled trajectory. The BFGS optimization routine is used to select the neural network parameters subject to the loss function. The convergence criteria was arbitrarily set to obtaining a loss of less than $10^{-5}$.

\subsection{FitzHugh-Nagumo Wavetrain} \label{app:fhn}

The simulation of Section \ref{sec:fhn} was performed via third-order finite differences in space and a fourth-order Runge-Kutta integrator in time. The model parameters used are listed in Table \ref{tab:fhn}. The periodic 1-D domain is $x \in [0,500)$ with 500 grid points. The initial condition is $u(x,t=0) = \frac{1}{2}\left(1 + \sin\left( \frac{10 \pi}{500} x \right) \right)$ and $v(x,t=0) = \frac{1}{2}\left(1 + \cos\left( \frac{10 \pi}{500} x \right) \right)$. The system is integrated to time $t = 1000$. The snapshot matrix contains 401 snapshots (including the initial condition).

\begin{table}
	\caption{FHN Simulation Parameters}
	\label{tab:fhn}
	\centering
	\begin{tabular}{ccccc}
	\hline
	\hline
	Parameter & $a$ & $b$ & $\nu$ & $\epsilon$ \\ \hline
	Value  & -0.1 & 0.3 & 1 & 0.05 \\

	\hline
	\hline
	\end{tabular}
\end{table}

The neural ODE surrogate model uses an automatic stiffness-detecting ODE solver that selects between a fifth-order Runge-Kutta integrator and a second-order Rosenbrock integrator. The neural network is constructed with the knowledge that $U' = W$: 
\begin{equation}
\begin{cases}
    U' = W \\
    V' = {NN}_1 \left(U,V,W\right) \\
    W' = {NN}_2 \left(U,V,W\right) .
\end{cases}
\end{equation}
The neural network is comprised of an input layer of dimension three, one fully-connected hidden layer of size 3 (sigmoid-activated), and an output layer of dimension two (linear). The last snapshot of the simulation data is used as the training data. To avoid issues in training associated with local minimia and periodic trajectories, the loss function is constructed to contain multiple trajectories whose initial conditions are from regularly spaced along the simulation data:
\begin{equation}
    \mathcal{L} = \sum_{i=1}^{N/k} {\left( \mathbf{y}(x_{ki} + \Delta \xi) - \ODESolve(\mathbf{y}(x_{ki}),NN_\theta)\Bigr\rvert_{\xi = \Delta \xi} \right)}^2 ,
\end{equation}
where $N$ is the total number of data points in the trajectory, $k$ is the degree of sub-sampling (every $k$-th data point is used to seed an ODESolve), and $\Delta \xi$ sets where along the trajectories the ODESolve is evaluated. For this example, $k = 4$ and $\Delta \xi = 5$. The BFGS optimization routine is used to select the neural network parameters subject to the loss function. The convergence criteria is arbitrarily set to obtaining a loss less than $10^{-5}$.

\bibliographystyle{ieeetran}
\bibliography{tw}

\end{document}